\documentclass[aps, prd, twocolumn, preprintnumbers, superscriptaddress, nofootinbib, floatfix]{revtex4-1}

\usepackage{amsmath}
\usepackage{bm}
\usepackage{amsfonts}
\usepackage{graphicx}
\usepackage{epstopdf}
\usepackage{hyperref}
\usepackage{array}
\usepackage{soul}
\usepackage{color}

\newcommand{\exclude}[1]{}

\enlargethispage{\baselineskip}

\hypersetup{
     colorlinks   = true,
     citecolor    = blue
}

\newcommand{\TRH}{T_{\rm RH}}
\newcommand{\MP}{M_{\rm Pl}}

\renewcommand\({\left(}
\renewcommand\){\right)}
\renewcommand\[{\left[}
\renewcommand\]{\right]}
\newcommand{\be}{\begin{equation}}
\newcommand{\ee}{\end{equation}}
\newcommand{\bea}{\begin{eqnarray}}
\newcommand{\eea}{\end{eqnarray}}

\begin{document}

\title{\bf Probing the Early Universe with Axion Physics and Gravitational Waves}

\newcommand{\FIRSTAFF}{\affiliation{Department of Physics and Astronomy, Uppsala University, L\"agerhyddsv\"agen 1, 75120 Uppsala, Sweden}}
\newcommand{\SECONDAFF}{\affiliation{Nordita, KTH Royal Institute of Technology and Stockholm University, Roslagstullsbacken 23, 10691 Stockholm, Sweden}}
\author{Nicklas Ramberg}\email[Electronic address: ]{nicklasramberg1993@gmail.com}\FIRSTAFF
\author{Luca Visinelli}\email[Electronic address: ]{luca.visinelli@physics.uu.se} \FIRSTAFF \SECONDAFF
\date{\today}
\begin{abstract}
	We show results for the expected reach of the network of experiments that is being set up globally with the aim of detecting the ``invisible'' axion, in light of a non-standard thermal history of the universe. Assuming that the axion is the dark matter, we discuss the reach of a successful detection by a given experimental setup in a particular axion mass window for different modifications of the cosmological background before primordial nucleosynthesis occurred. Results are presented both in the case where the present energy budget in cold axions is produced through the vacuum realignment mechanism alone, or in the case in which axionic strings also provide with additional contributions to the axion energy density. We also show that in some cosmological models, the spectrum of gravitational waves from the axionic string network would be within reach of the future network of detectors like LISA and DECIGO-BBO. We conclude that some scenarios describing the early universe can be probed jointly by the experimental efforts on axion detection and by gravity wave multi-messenger astronomy.
\end{abstract}
\maketitle

\section{Introduction}

The $\Lambda$-Cold Dark Matter or $\Lambda$CDM paradigm is a concordance cosmological model that best reconciles various, apparently inconsistent, observed properties of the early universe~\cite{Cole:2005sx, 2011ApJ, 2011ApJ743, 2011MNRAS, 2013ApJS, Anderson:2013zyy, Ade:2015tva, Ade:2015xua}. In the $\Lambda$CDM model, most of the energy content of the universe is stored in Dark Matter (DM), around 27\%, and Dark Energy (DE), around 68\%, the remaining 5\% comprising the ordinary ``baryonic'' matter. Energy components into radiation and neutrinos are at present negligible, although they played a fundamental role in driving the evolution of the universe at earlier stages. The $\Lambda$CDM model takes general relativity as the correct theory of gravity acting on cosmological scales. In the era of precision cosmology, much is known about the content and the evolution of the metric in the early universe, with our knowledge of the history of the universe extending much before the epoch of recombination at which the Cosmic Microwave Background Radiation (CMBR) decouples from baryons. In this view, the oldest relics we can use to probe the early universe are the very fractional abundances of light elements that originated during Big-Bang Nucleosynthesis (BBN), one of the most important predictions obtained from jointly considering cosmology and particle physics (see Refs.~\cite{Tytler:2000qf, 2016:Cyburt} for reviews on the topic). At temperatures of the order of approximately $0.1\,$MeV, light elements like deuterium and Helium isotopes synthesis and are not dissociated by CMBR scattering. Any change in this picture coming from additional new physics would alter the relative fractions of light elements forming at BBN, so that it is usually assumed that the $\Lambda$CDM model holds up to a temperature $T_{\rm BBN} \approx 5$\,MeV that ensures successful BBN without spoiling its results~\cite{Kawasaki:1999na, Kawasaki:2000en, Hannestad:2004px, ichikawa:2005vw, DeBernardis:2008zz}.

The detection of a new relic that carries information of the pre-BBN universe would further push our knowledge to probe the cosmo at temperatures much larger than $T_{\rm BBN}$. Such a relic could be in the form of primordial gravitational waves, a dark matter particle, or any new particle forming an unknown ``hidden'' sector. It is not at all guaranteed that the pre-BBN epoch had always been described by the $\Lambda$CDM model, since new particles from hidden sectors could have come to dominate the expansion rate of the universe for some period before either decaying or redshift away. The evolution of the post-inflationary universe down to the temperature $T_{\rm BBN}$ is still unknown, and motivates considering non-standard cosmologies (NSCs) that alter the $\Lambda$CDM paradigm due to the lack of additional information.

Such a modification to the $\Lambda$CDM model at temperatures larger than $T_{\rm BBN}$ would lead to several testable consequences in various phenomenological aspects. In fact, the presence of an early NSC would alter the properties at which a Weakly Interacting Massive Particle (WIMP) free
zes out~\cite{Gelmini:2006mr, Gelmini:2008sh, Erickcek:2011us, Waldstein:2016blt, Visinelli:2017qga} and decouples kinetically from the primordial plasma~\cite{Visinelli:2015eka, Waldstein:2017wps}, and open the pathway to testable modifications to the standard leptogenesis models~\cite{Dutta:2018zkg}. In general, if a new relic is ever discovered, the properties of its phase-space distribution would carry crucial information on the pre-BBN era.

An early NSC period would also alter the property of the ``invisible'' QCD axion~\cite{Weinberg:1977ma, Wilczek:1977pj}, a light Goldstone boson arising within the solution to the strong CP problem proposed by Peccei and Quinn (PQ)~\cite{Peccei:1977ur, Peccei:1977hh}, and makes up for a plausible dark matter candidate~\cite{Abbott:1982af, Dine:1982ah, Preskill:1982cy}. In the standard cosmological scenario, refined cosmological simulations yield a narrow range in which the QCD axion would be the CDM particle, with an axion mass in the range $m_A = \mathcal{O}\(10\){\rm \,\mu eV}$~\cite{Visinelli:2009zm, Visinelli:2014twa}. These results have been recently proven by refined cosmological simulations~\cite{Klaer:2017qhr, Klaer:2017ond, Gorghetto:2018myk, Vaquero:2018tib}. Considering a NSC model in the pre-BBN epoch considerably widens the mass window of CDM axions~\cite{Visinelli:2009kt, Visinelli:2018wza}, alter the properties carried by non-relativistic axions like their present energy density and velocity distribution~\cite{Visinelli:2009kt, Visinelli:2017imh, Visinelli:2018wza, Nelson:2018via}, and alter the energy content of hot axions~\cite{Grin:2007yg}. These possibilities have been proven interesting for a number of experiments that are planned to explored the parameter space of the QCD axion and other axion-like particles away from the preferred region, see Refs.~\cite{Raffelt:1995ym, Raffelt:2006rj, Sikivie:2006ni, Kim:2008hd, Wantz:2009it, Kawasaki:2013ae, Marsh:2015xka, Kim:2017yqo, Irastorza:2018dyq} for reviews.

An early NSC would also alter the evolution and the decay of a network of global strings that generates after the spontaneous braking of a global symmetry. Since the axion field originates from the spontaneous breaking of a global symmetry, it is expected that a network of global strings is formed along with it~\cite{Vilenkin:1981kz, Vilenkin:1984ib}, lasting until the moment $t_{\rm osc}$ at which the axion field acquires a mass and begins to oscillate about its minimum~\cite{Davis:1986xc}. A large literature on strings is available, due to the fact that such objects have been considered a candidate for seeding structure formations and play the role of the dark matter. A network of axionic strings is produced at the PQ phase transition through the Kibble mechanism~\cite{Kibble:1976}, with a non-trivial dynamics that leads an individual ``long'' strings to form kinks and cusps, or two long strings to intersect and form smaller wiggling loops. These new features would release energy by oscillating and releasing the massless Goldstone bosons associated with the broken global symmetry, like the axion, or other radiative modes if the Goldstone boson is massive. The decay of the axionic string network would sensibly contribute to the present abundance of cold axions~\cite{Vilenkin:1982ks, Harari:1987ht, Vilenkin:2000jqa}, as it has been recently confirmed using refined cosmological simulations~\cite{Hiramatsu:2010yn, Hiramatsu:2012sc, Hiramatsu:2012gg, Klaer:2017qhr, Klaer:2017ond, Gorghetto:2018myk, Vaquero:2018tib}, with earlier numerical work on the subject found in Refs.~\cite{Battye:1993jv, Battye:1994au, Yamaguchi:1998gx, Yamaguchi:1999dy}.

A different decay channel of the string network is through the modal emission of gravitational waves (GWs), generated from radiating string loops if the Goldstone bosons like the axion are massive. Gravitational waves are produced by various mechanisms like the oscillation of strings in a network, kinks and cusps of long strings, or the emission from the oscillation of loops which dissipate and are replaced by new loops forming from the chopping of long strings~\cite{Battye:1997ji, Damour:2000wa}, and have recently received substantial attention due to their possible appearance in the upcoming arrays of detectors~\cite{BlancoPillado:2011dq, Blanco-Pillado:2013qja, Blanco-Pillado:2017oxo}. The inclusion of an early NSC stage would lead to additional features in the relic GW spectrum which could be detectable in the upcoming generation of dedicated experiments~\cite{Cui:2018rwi} and would in general differ from the features expected from other mechanisms like GW from inflation~\cite{Starobinsky:1979ty, Sahni:1990tx}, from the displacement of the Higgs field~\cite{Amin:2019qrx}, or from extensions of the theory of general relativity~\cite{Novikov:2016hrc, Novikov:2016fzd}. There has also been recent interest in the implication of an early quintessential NSC on the spectrum of GW from cosmic strings~\cite{Kamada:2014qja, Bettoni:2018pbl}, while earlier discussions on the non-thermal production from the decay of cosmic strings of different dark matter candidates has been sketched in Refs.~\cite{Jeannerot:1999yn, Matsuda:2005fb, Cui:2008bd}. A review on the subject is available in Ref.~\cite{Maggiore:1999vm}.

In this paper we explore the consequences of an early NSC on the energy density and the distribution of axions and gravitational waves emitted from axionic strings. We thus consider the cold axion as the main ingredient of the dark matter content of the $\Lambda$CDM model~\footnote{For a review of dark matter model see Refs.~\cite{Bertone:2004pz, Garrett:2010hd, Gelmini:2015zpa}. Mixed WIMP-axion models have been considered in Refs.~\cite{Bae:2014efa, Bae:2015rra, Baum:2016oow}. For an explanation of the rotation curves of galaxies in terms of modified gravity see Refs.~\cite{Famaey:2005fd, Milgrom:2012xw, Vagnozzi:2017ilo}.}. The earlier work on the subject presented in Ref.~\cite{Visinelli:2009kt} has been considerably extended by including a wider variety of NSCs, while the treatment on the axionic strings has been carefully taken into account by detailing the evolution of the string network and the impact on the energy density of the radiated axions. We have implemented a numerical analysis that solves for the set of coupled kinetic equations describing the evolution of the energy densities of radiation and the new fluid during the NSC period, following the treatment in Refs.~\cite{Giudice:2000dp, Giudice:2000ex, Giudice:2001ep, Visinelli:2014qla, Visinelli:2016rhn, Freese:2017ace}. In Sec.~\ref{sec:axion_misalignment}, we study the dependence of the abundance of axion CDM produced through the vacuum realignment mechanism by solving the equation of motion for the axion field numerically and obtaining the value of the axion mass with a bisection method. In Sec.~\ref{sec:axion_strings} we add the contribution originated from the decay of the string network loops to the axion energy density from the decay of the string network, which is also described by coupled kinetic equations, following the discussion in the standard cosmological scenario~\cite{Battye:1994au, Vilenkin:2000jqa}. Assuming that the cold axion is the CDM particle, we show the potential reaches of current future axion searches in the parameter space that describes the modified cosmological model and comprises the equation of state for the new fluid or set of fluids, here $w_{\rm eff}$, and the reheating temperature at which the cosmology transitions to the $\Lambda$CDM model, here $\TRH$. The two additional parameters come from the axion theory, and are given by the axion mass $m_A$ and the value of the misalignment angle at the moment of the spontaneous symmetry breaking $\theta_i$.

We assume that the network of axionic strings decays predominantly into axions contributing to the present CDM density, with a subdominant spectrum of GWs which we have derived in Sec~\ref{Gravitational Waves from Axionic String Loops}. While in general we obtain the well-known result that the spectrum of GWs is negligible, for some particular choices of the parameter space corresponding to an early matter-dominated (MD) stage with a low reheating temperature we obtain a potential detection in the upcoming network of detectors, as we show in Fig.~\ref{fig:GWFigure} below.

\section{Background Evolution in a NSC Scenario}

In its simplest picture, the post-inflationary Universe has been dominated by a thermal bath of relativistic particles produced by a successful reheating mechanism, down to the temperature of matter-radiation equality. However, the cosmology realised in the early Universe might have differed from this view in many aspects. Besides a modification of General Relativity~\footnote{E. Nardi, M. Giannotti, L. Visinelli, in preparation}, various models that extend the particle content of the Standard Model of particle physics predict the existence of additional particles whose energy density might have dominated the expansion rate at some moment prior Big-Bang Nucleosynthesis (BBN). We actually know that the BBN model reproduces the abundance of various light elements with extreme precision and we do not want to spoil these results. In order for a successful nucleosynthesis to take place, it is required that the history of the universe be the standard radiation-dominated cosmology at temperatures $T_{\rm BBN} \sim 5{\rm \, MeV}$~\cite{Kawasaki:1999na, Kawasaki:2000en, Hannestad:2004px, ichikawa:2005vw, DeBernardis:2008zz}.

Generally speaking, the pre-BBN evolution of the universe in a NSC is altered by letting the equation of state $w_{\rm eff}$ of the fluid that dominates the expansion of the background differ from that of a relativistic content $w_{\rm rad} = 1/3$. In a particle physics model, this can be achieved in various different ways, for example by considering the early domination of a scalar field $\Phi$ that oscillates in the potential $\propto \Phi^{2n}$ and for which the effective equation of state is $w_{\rm eff} = (n-1)/(n+1)$~\cite{Turner:1983he}. For a quadratic potential $n=1$~\footnote{Fragmentation actually lowers the possible outcome of the effective equation of state~\cite{Lozanov:2016hid, Lozanov:2017hjm}, which is found to be $w_{\rm eff} = 0$ for $n=1$, while it reaches $w_{\rm eff} = 1/3$ for $n > 1$ after a relaxation period}, we obtain the early matter-domination period $w_{\rm eff} = 0$ discussed in previous literature~\cite{Dine:1982ah, Steinhardt:1983ia, Turner:1983he, Scherrer:1984fd}. A constant potential $n=0$, in which the field does not roll, leads to an inflationary period with $w_{\rm eff} = -1$, while in an extremely steep potential $n\to +\infty$ the potential energy density is suddenly converted into the kinetic energy of a ``fast-rolling'' field in its kination stage~\cite{Barrow:1982, Ford:1986sy, Spokoiny:1993kt, Joyce:1996cp, Salati:2002md, Profumo:2003hq}. A repulsive ``ultra-light'' scalar boson would also give rise to early kination-, radiation- and matter-like cosmological evolutions~\cite{Li:2016mmc}. If the particle theory that justifies the NSC scenario is described by a scalar field in a potential $V(\Phi)$ with equation of state
\begin{equation}
    w_{\rm eff} = \bigg\langle\frac{\dot\Phi^2 - 2V(\Phi)}{\dot\Phi^2+2V(\Phi)}\bigg\rangle,
    \label{eq:effective_omega}
\end{equation}
where brackets indicate time average, then we are constrained to $-1\leq w_{\rm eff}\leq 1$ unless tachyon forms of the kinetic term are considered. The possibility for tachyon fields are not considered here. The background modelling by a scalar field does not capture other important NSC given by the early domination of topological defects like cosmic strings $w_{\rm str} = -1/3$ or domain walls $w_{\rm wall} = -2/3$.

The background evolution could be realised by a mixture of different components, each with energy density $\rho_i(a)$ and effective equation of state $w_i$, so that an effective equation of state takes place with
\begin{equation}
    w_{\rm eff}(a) = \frac{\sum_i\,w_i\,\rho_i(a)}{\sum_i\rho_i(a)}.
\end{equation}
For this reason, in the following we consider an effective equation of state for $w_{eff}$ ranging over the interval $\[-1; 1\]$, although we do not take into account the possible temporal evolution of such an equation of state.

We label the effective fluid as $\Phi$, and we consider its equation of state $w_{\rm eff}$, bearing in mind that the NSC could have been described by a mixture of components, some of which might not have been particle fields. The set of kinetic equations describing the conversion of the energy density of the $\Phi$ field into radiation is 
\begin{eqnarray}
    \dot\rho_\Phi + 3(1+w_{\rm eff})H\rho_\Phi &=& -Q, \label{eq:Kinetic_psi}\\
    \dot\rho_R + 4H\rho_R &=& Q, \label{eq:Kinetic_Rad}
\end{eqnarray}
where a dot indicates a derivation with respect to cosmic time $t$ and $Q$ is a source term that ensures the decay of the $\Phi$ field. Notice that, for $Q=0$, the solution to Eq.~\eqref{eq:Kinetic_psi} is
\begin{equation}
    \rho_\Phi \propto a^{-3(1+w_{\rm eff})}, \quad\hbox {for $Q = 0$}.
    \label{eq:sol_psi_noQ}
\end{equation}
Thus, for $w_{\rm eff} > 1/3$, the additional energy density component redshifts away faster than radiation, so that a ``$\Phi$-radiation'' equality $a_{\Phi R}$ is reached. If the temperature at which such a transition occurs is high enough not to spoil nucleosynthesis, no extra decay term is needed, as it is the case for the kination model for which $w_{\rm eff} = 1$. Here, we consider a non-zero source term of the form $Q = \Gamma\rho_\Phi$, where the parameter $\Gamma$ controls the rate of decay. With this decay rate, the expression in Eq.~\eqref{eq:Kinetic_psi} can be evaluated exactly to yield
\begin{equation}
    \rho_\Phi = \rho_{\Phi 0}\, a^{-3(1+w_{\rm eff})}\,\exp\left(-\Gamma t\right), \quad\hbox {for $Q = \Gamma\rho_{\Phi}$},
\end{equation}
where $\rho_{\Phi 0}$ is the initial value of the $\Phi$ field. Well inside the $\Phi$-domination regime, when $t \ll \Gamma^{-1}$, the source term can be neglected and the solution in Eq.~\eqref{eq:sol_psi_noQ} is recovered. In this regime, assuming that the $\Phi$ field guides the expansion rate of the Universe at such an early time, the Hubble rate is given by
\begin{equation}
    H =\sqrt{\frac{\rho_\Phi}{3\MP^2}} \propto a^{-\frac{3}{2}(1+w_{\rm eff})},
    \label{eq:Hubble_modified}
\end{equation}
where $\MP$ is the reduced Planck mass. Notice that this relation predicts the behaviour $a(t) \propto t^\beta$, where for simplicity we have introduced $\beta = 2/3(1+w_{\rm eff})$ so that $a(t) \propto t^\beta$ well inside the NSC period. The energy density in radiation is given by the particular solution to Eq.~\eqref{eq:Kinetic_Rad} with source term $Q$, which we write using the scale factor as the independent variable as
\be
    \frac{1}{a^3}\frac{d}{da}\(a^4\rho_R\) = \frac{\Gamma}{H}\rho_\phi = \sqrt{3}\MP\Gamma \rho_\phi^{1/2},
    \label{eq:Kinetic_Rad1}
\ee
with solution $\rho_R \propto a^{-\frac{1 + 3w_{\rm eff}}{2}}$. For a decaying $\Phi$ field, the Hubble rate is then given by $H\propto \rho_R$~\cite{Visinelli:2018wza}. Although the Universe is dominated by the $\Phi$ field, we can nevertheless define the temperature associated to the radiation bath as $T \propto \rho_R^{1/4}$, which leads to a relation between the Hubble rate and temperature of the form $H \propto T^4$, which is the same proportionality obtained in previous literature for the case $w_{\rm eff} = 0$~\cite{Giudice:2000ex, Giudice:2001ep}. However, we have just shown that such a proportionality holds for any equation of state for a decaying field, including the case of a decaying kination field~\cite{Visinelli:2018wza}.

Although the solution to the background evolution is in principle dependent on the choice of the initial condition, we obtain that the dynamics of the axion field is not altered for sufficiently large values of $\rho_{\Phi 0}$ since, as we discuss below in Sec.~\ref{sec:axion_misalignment}, the axion field is frozen down to temperatures of the order of the GeV.

\section{Axion Physics}

The sector describing the strong interactions of the Standard Model possesses a non-trivial vacuum that allows for a violation of the CP symmetry. Jointly with effects generated by weak interactions, the amount of such a violation is constrained by the non-detection of a neutron electric dipole moment to be $\theta \lesssim 10^{-10}$. The strong-CP problem is formulated as the question why the parameter $\theta$ is realised with such a small value, while a priori it could realise any value within $[-\pi, \pi]$~\cite{Belavin:1975fg, tHooft:1976rip, Jackiw:1976pf, Callan:1976je}. An elegant solution to the problem is exemplified by the Peccei-Quinn (PQ) model~\cite{Peccei:1977hh, Peccei:1977ur}, in which a new global chiral U(1) PQ symmetry is introduced. The PQ symmetry is spontaneously broken when the PQ complex scalar field rolls down the minimum of a Mexican hat potential
\begin{equation}
    V(\Phi) = \lambda\,\left(\left|\Phi\right|^2 - \frac{f_A^2}{2}\right)^2,
    \label{eq:PQpotential}
\end{equation}
where the complex scalar field can be decomposed as $\Phi = (f_A/\sqrt{2})\exp\left(i\phi/f_A\right)$, where the pseudo-scalar field $\phi$ is the axion and $f_A$ is a yet unknown energy scale known as the axion decay constant. The quantum chromodynamics (QCD) axion is the pseudo Nambu-Goldstone boson associated with the spontaneous breaking of the PQ symmetry~\cite{Weinberg:1977ma, Wilczek:1977pj}, which is an anomalous symmetry since it is explicitly broken by QCD effects. The mass of the axion arises from small QCD effects and depends on temperature so that $m_A^2(T) = \chi(T)/f_A^2$, where the susceptibility $\chi(T)$ is obtained from lattice computations~\cite{Borsanyi:2016ksw}. At zero temperature, the axion mass is~\cite{Weinberg:1977ma}
\begin{equation}
	m_A = \frac{\sqrt{z}}{1+z}\frac{\Lambda_A^2}{f_A/N} = 6.2 {\rm \mu eV}\left( \frac{10^{12}{\rm GeV}}{f_A/N}\right),
	\label{eq:axionmass}
\end{equation}
where $z$ is the ratio of the masses of the up and down quarks while $\Lambda_A = \chi^{1/4}(0) = 75.5\,$MeV. The integer $N$ represents the $U(1)_{PQ}$ color anomaly index, which we set $N=1$.

Active searches for axions are undergoing, using the fact that an anomalous axion-photon coupling arises with strength set by the scale $f_A^{-1}$, leading to possible ``axion electrodynamics'' effects~\cite{Wilczek:1987mv, Krasnikov:1996bm, Li:2009tca, visinelli:2013fia, Tercas:2018gxv, Visinelli:2018zif} which has been translated into searches with microwave cavity detectors~\cite{Sikivie:1983ip, Sikivie:1985yu}. In a cavity, the axion is interacting with the magnetic field and resonantly converts into a quasi-monochromatic signal which is picked up by an antenna and then amplified to the audible range~\cite{graham2015experimental}. Such a technique has been fruitfully translated into a vigorous laboratory detection program by some collaborations like ADMX~\cite{PhysRevLett.104.041301, asztalos2011design, PhysRevD.74.012006, graham2015experimental, Stern:2016bbw}, YLW~\cite{Brubaker:2016ktl} and KLASH~\cite{Alesini:2017ifp}. A different technique that implements a dielectric haloscope has been implemented by the MADMAX collaboration~\cite{TheMADMAXWorkingGroup:2016hpc, Majorovits:2016yvk}, while the ABRACADABRA experiment exploits the coupling of the dark matter axion to a static magnetic field by probing the oscillating magnetic field induced by the particle~\cite{Kahn:2016aff, Ouellet:2018beu}. For thorough reviews on the topic of the QCD axion, we refer the reader to Refs.~\cite{Raffelt:1995ym, Raffelt:2006rj, Sikivie:2006ni, Kim:2008hd, Wantz:2009it, Kawasaki:2013ae, Marsh:2015xka, Kim:2017yqo}, while experimental searches have been reviewed in Refs.~\cite{Irastorza:2018dyq}. Other methods of detection rely on the interaction with compact structures such as axion miniclusters~\cite{Hogan:1988mp, Kolb:1993zz, Kolb:1993hw, Kolb:1994fi, Sakharov:1996xg, Enander:2017ogx, Vaquero:2018tib, Visinelli:2018wza, Nelson:2018via} and axion stars~\cite{Kaup:1968zz, Ruffini:1969qy, Das_1963, Feinblum:1968nwc, Teixeira:1975ad, Colpi:1986ye, Seidel:1991zh, Tkachev:1991ka, Chavanis:2011, Braaten:2015eeu, Eby:2015hyx, Eby:2016cnq, Levkov:2016rkk, Helfer:2016ljl, Braaten:2016dlp, Braaten:2016kzc, Bai:2016wpg, Eby:2017xaw, Desjacques:2017fmf, Visinelli:2017ooc, Chavanis:2017loo, Krippendorf:2018tei}. Dark matter axions in NSC have been treated extensively in Refs.~\cite{Visinelli:2009kt, Visinelli:2017imh, Visinelli:2018wza, Draper:2018tmh, Nelson:2018via, Blinov:2019rhb}.

\section{Results for the vacuum realignment mechanism in NSC} \label{sec:axion_misalignment}

The QCD axion could serve as a particle explanation for the dark matter~\cite{Preskill:1982cy, Abbott:1982af, Dine:1982ah}, with an energy density that generates non-thermally after the PQ symmetry breaking occurring at a yet unknown energy scale $f_A$. In more details, taking the variation of the action for the weakly coupled PQ field with the potential in Eq.~\eqref{eq:PQpotential}, the equation of motion for the axion misalignment angle $\theta = \phi/f_A$ reads
\begin{equation}
    \ddot\theta + 3H\dot\theta + m_A^2(T)\sin\theta = 0,
    \label{eq:axioneqmotion}
\end{equation}
where a dot indicates a derivation with respect to cosmic time $t$. The mass of the QCD axion depends on temperature because of the interaction with QCD instantons~\cite{Gross:53.43, Fox:2004kb}. The dependence of the axion susceptibility $\chi(T) \equiv f_A^2m_A^2(T)$ with temperature has been computed with semi-analytical methods~\cite{Turner:1986, Bae:2008ue, Wantz:2009it, diCortona:2015ldu} and recently using refined lattice simulations~\cite{Borsanyi:2015cka, Borsanyi:2016ksw, Petreczky:2016vrs}. Here, we use the numerical result obtained in Ref.~\cite{Borsanyi:2016ksw} to parametrise the temperature-dependence of the susceptibility.

Given an initial value of the axion angle $\theta_i$, drawn randomly from the uniform distribution $[-\pi, \pi]$ when the PQ phase transition occurs, the solution to Eq.~\eqref{eq:axioneqmotion} is a constant value of $\theta = \theta_i$ as long as the Hubble friction is much larger than the axion mass. The axion field starts to oscillate when the condition $3H(T_{\rm osc}) \approx m_A(T_{\rm osc})$ is met, from which moment the number of axions in a comoving volume is fixed and the axion energy density scales as $\rho_A \propto a^{-3}$. Here, we solve numerically the Sine-Gordon Eq.~\eqref{eq:axioneqmotion} describing the axion field, using as the independent coordinate the rescaled temperature $\tau = T/T_{\rm osc}$,
\begin{eqnarray}
    \theta'' &+& F(\tau)\frac{\theta'}{\tau} + \omega^2(\tau)\sin\theta = 0,\\
    F(\tau) &\equiv& \left(\frac{d\tau}{dt}\right)^{-2}\left[3H(\tau)\frac{d\tau}{dt} + \frac{d^2\tau}{dt^2}\tau\right],\\
    \omega^2(\tau) &\equiv& \left(\frac{d\tau}{dt}\right)^{-2}\frac{m^2}{T_{\rm osc}^2}.
\end{eqnarray}
we obtain the energy density at some temperature $T_* < T_{\rm osc}$ at which the number of axions in a comoving volume does not change, so that we have attained the conservation of the quantity $N_A^* = \rho_A(T_*) a^3(T_*)/m_A(T_*)$. The present fractional axion energy density is then
\begin{equation}
    \Omega_A = \frac{\rho_A(T^*)}{\rho_c(t_0)}\frac{m_A(T_0)}{m_A(T_*)}\left(\frac{a_*}{a_{\rm RH}}\right)^3\,\frac{g_S(T_0)}{g_S(T_{\rm RH})}\frac{T_0^3}{T_{\rm RH}^3},
\end{equation}
where we have assumed the conservation of the entropy density from the end of the NSC at $T_{\rm RH}$ and where the fraction $a_*/a_{\rm RH}$ is computed numerically from the solution to the background. In the last expression, $\rho_c(t) = 3\MP^2H(t)^2$ defines the critical energy density at time $t$, so that $\rho_c(t_0)$ is the present critical density. For a given initial condition $\theta_i$, we loop our procedure over the value of the axion mass to obtain the value that yields the observed CDM abundance $\Omega_{\rm CDM}h^2 \sim 0.12$. The Hubble rate used in the numerical computation describes the NSC scenario and it is obtained by solving the set of kinetic Eqs.~\eqref{eq:Kinetic_psi}-\eqref{eq:Kinetic_Rad} for the energy densities $\rho_\Phi$ and $\rho_R$.

\begin{figure}[tb]
\begin{center}
	\includegraphics[width=\linewidth]{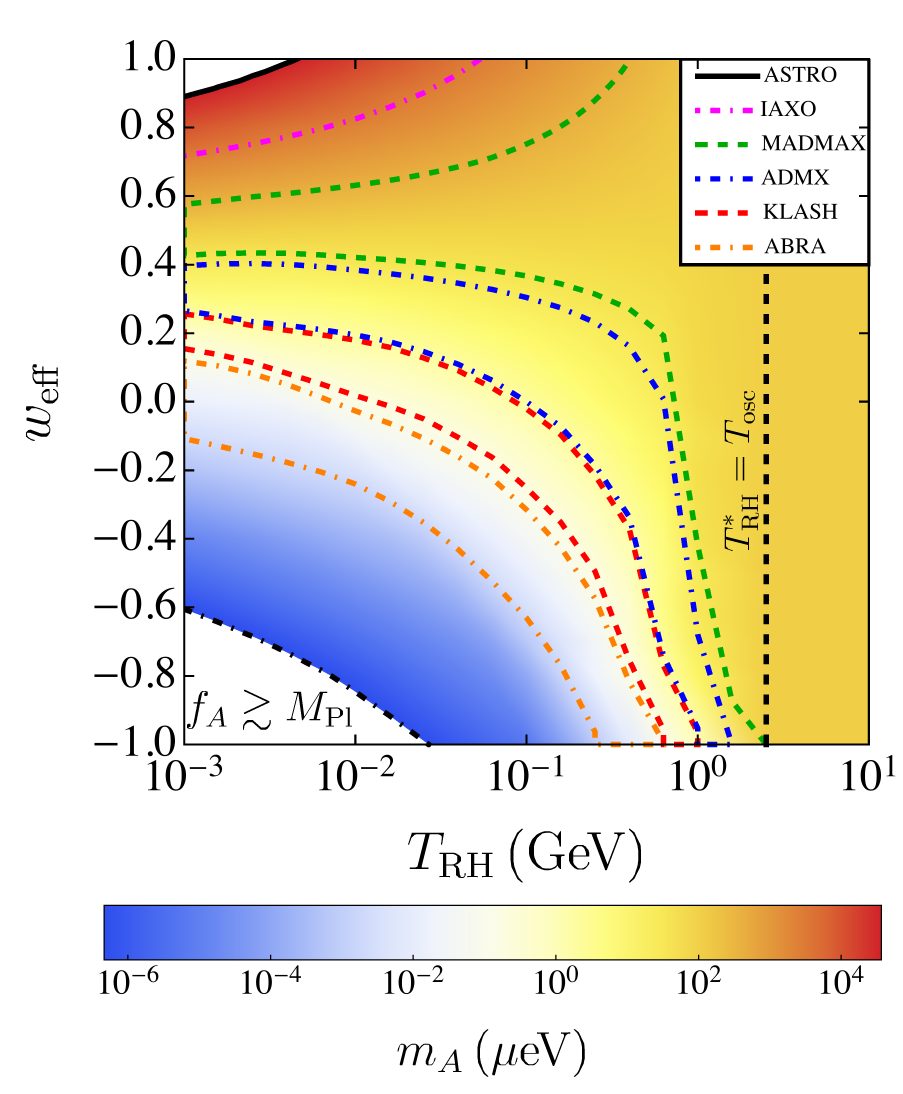}
	\caption{The value of the axion mass (in ${\rm \mu eV})$ that yields the observed dark matter abundance for different cosmological scenarios. The vertical axis reports the effective equation of state $w_{\rm eff}$, while the horizontal axis is the reheat temperature. The dashed vertical black line marks the reheat temperature $\TRH^*$ that matches the temperature at which the axion acquires a mass in the standard scenario, $T_{\rm RH}^* = T_{\rm osc}$. We also show the sensitivity that is expected to be reached by ABRACADABRA (``ABRA'', orange dot-dashed line), KLASH (red dashed line), ADMX (blue dot-dashed line), MADMAX (green dashed line), and IAXO (magenta dot-dashed line). The black dashed line marks the region beyond which the axion decay constant violates the weak gravity conjecture, $f_A \gtrsim \MP$, while the black solid line ``ASTRO'' gives the bound on $m_A \gtrsim 15{\rm meV}$ excluded by astrophysical considerations~\cite{RAFFELT1986402, Raffelt2008, Viaux:2013lha, Giannotti:2017hny}.}
	\label{fig:axionmass_nostrings}
\end{center}
\end{figure}

In Fig.~\ref{fig:axionmass_nostrings} we show the value of the axion mass (in units of ${\rm \mu eV}$) that is required to attain the totality of the CDM for given reheating temperature and effective equation of state of the additional energy component. We have fixed the value of the misalignment angle $\theta_i = \pi/\sqrt{3}$. In Fig.~\ref{fig:axionmass_nostrings}, we only shows the contribution to the axion energy density coming from the misalignment mechanism, while the additional contribution from axionic strings has been shown in Fig.~\ref{fig:RecapFigureStrings} below. For each choice of the parameters, the corresponding value of $m_A$ is the smallest value of the axion mass attained in the theory, since smaller values of the would yield to a larger axion energy density than what is observed in dark matter. Higher values of the axion mass are possible, although the corresponding energy density would not lead to the totality of the dark matter. The dashed vertical black line marks the reheat temperature $T_{\rm RH}^*$ for which $T_{\rm RH}^* = T_{\rm osc}$, which can be approximated as the solution to the expression $m_A(T_{\rm RH}^*) = 3H(T_{\rm RH}^*)$. An axion field would be in a frozen configuration $\theta = \theta_i$ for the whole duration of the NSC with $T_{\rm RH} > T_{\rm RH}^*$ and would acquire a mass only at a lower temperature $T_{\rm osc}$. In this scenario, the axion mass is not sensitive to the details of the NSC. As we will discuss later in Sec.~\ref{sec:axion_strings}, the situation in the presence of a string network is different since axions are released through the decay of string loop at all time prior the demise of the network.

In Fig.~\ref{fig:axionmass_nostrings} we have indicated the reach that is forecast by various experiments for the QCD axion in the KSVZ theory. In the ABRACADABRA proposal~\cite{Kahn:2016aff}, a large portion of the axion parameter space is possible within reach on the next decay, with a prototype~\cite{Ouellet:2018beu} that has started to collect data and place experimental bounds. KLASH~\cite{Alesini:2017ifp} is going to look for axions in the mass range $(0.2-2){\rm \,\mu eV}$, ADMX~\cite{Stern:2016bbw} is expected to extend the search in the mass range $(3-25){\rm \,\mu eV}$, MADMAX~\cite{TheMADMAXWorkingGroup:2016hpc} is going to search for axions in the mass range $(40-400){\rm \,\mu eV}$, and IAXO~\cite{Vogel:2013bta, Giannotti:2016drd, Armengaud:2019uso} will be sensitive to the axion-photon coupling $g_{a\gamma\gamma} \gtrsim 6\times 10^{-11}{\rm GeV}^{-1}$, which corresponds to a QCD axion of mass $ \gtrsim 10^4{\rm \mu eV}$. Contrary for the other experiments quoted in Fig.~\ref{fig:axionmass_nostrings}, for which the reach lies within a band in the axion mass window, the reach of the axion mass for IAXO corresponds to the portion of the parameter space above the magenta dot-dashed line in the figure. We have also included the region marked by $f_A \gtrsim \MP$ and bound by the black dot-dashed line that corresponds to the choices of the parameters for which the weak gravity conjecture is violated, $Gf_A^2 \lesssim 1$, and translates into the lower bound $m_A \gtrsim 4.6 \times 10^{-7}{\rm \,\mu eV}$ for the QCD axion. We have not taken into account here the potentially severe problem that arises when demanding to have an axion of quality~\cite{Giddings:1987cg, Kamionkowski:1992mf, Holman:1992us, Kallosh:1995hi, Lillard:2018fdt}. Notice that the lighter axions considered in the literature and known as ultra-light axions~\cite{Hu:2000ke} and realised in string-theory contexts~\cite{Svrcek:2006hf, Svrcek:2006yi, Arvanitaki:2009hb, Arvanitaki:2009fg, Higaki:2012ar, Cicoli:2012aq, Cicoli:2012sz, Higaki:2013lra, Stott:2017hvl, Visinelli:2018utg, Hoof:2018ieb} are not the QCD axion described here. The region of the parameter space already excluded by the non-observation of a faster cooling of astrophysical objects due to additional axionic channels are labeled as ``ASTRO''~\cite{RAFFELT1986402, Raffelt2008, Viaux:2013lha, Giannotti:2017hny}.

\section{Global strings contribution} \label{sec:axion_strings}

\subsection{The energy density of the string network}

The scenario described in Sec.~\ref{sec:axion_misalignment} holds as long as the PQ symmetry was broken during an inflationary period that washes out any topological defect~\cite{Linde:1987bx, Linde:1991km, Turner:1991, Wilczek:2004cr, Tegmark:2005dy, Hertzberg:2008wr, Freivogel:2008qc, Mack:2009hv, Visinelli:2009zm, Acharya:2010zx}, or if such topological defects can be neglected. If the Peccei-Quinn symmetry has broken after the period of inflation, or if the symmetry is temporarily restored after inflation to then break again, additional mechanisms contribute to the dark matter axion budget, including the decay of early topological defects~\cite{Kibble:1976, Kibble:1982dd}. In fact, the randomness of the initial condition of the axion field, jointly with the assumption of the continuity of the solution, leads to the prediction of the formation of an axionic string network~\cite{Davis:1985pt, Davis:1986xc}, topological configurations that result from the spontaneous breaking of a global symmetry~\cite{Vilenkin:1982ks} as non-trivial solution to the equation of motion of the PQ field in the potential in Eq.~\eqref{eq:PQpotential}. Besides axionic strings, other global string networks resulting from the breaking of a global symmetry have been studied in the context of realistic grand unified models~\cite{Vilenkin:1982ks}. It is expected that a network composed of approximately one axionic strings per Hubble volume is formed, releasing energy density during their evolution in the form of axions either by wiggles on long open strings or collapsing of closed strings. The cold portion of the spectrum of axions emitted from axionic strings might significantly contribute to the present energy density~\cite{Vilenkin:1981kz, Vilenkin:1982ks, Vilenkin:1986ku, Davis:1989nj, Battye:1993jv, Yamaguchi:1998gx, Hiramatsu:2010yu}. Due to their origin, axionic strings closely resemble global strings so that, in the standard cosmological scenario during radiation domination, the energy density per unit length of a single string diverges logarithmically, while the energy density of a network possesses a cutoff at a radius of the order of the horizon scale. In this model, axionic strings from broken global symmetries effectively dissipate by radiating a spectrum of axions. Here, we focus on the cosmological consequences of a string network, for which case global strings can be approximated by thin curves with a string core of size $\approx 1/f_A$ and a linear mass density $\mu_{\rm eff}$~\cite{Zeldovich:1978wj, Preskill:1979zi, Guth:1979bh, Vilenkin:1981kz, Vilenkin:1982ks}. 

The evolution of a string evolution is a complex physical system which demands describing structures on a wide range of scales, resulting in a significant computational demand when implemented numerically. Luckily, simulations in both matter- and radiation-dominated cosmologies show that the string network neatly divides into two distinct populations, a ``long'' string network with energy density $\rho_\infty$ and a set of small closed loops each of size $\ell$ and energy density $\rho_{\rm loop}$, so that the total energy density of the string network is $\rho_{\rm str} = \rho_\infty + \rho_{\rm loop}$. In the standard cosmology, it is expected that about 80\% of the energy density is in long strings, while the remaining 20\% is distributed into small loops spanning all lengths from the string core $\sim \pi/f_A$ to the horizon scale $\sim \pi/H(t)$~\cite{Gorghetto:2018myk}. The long string network is a Brownian random walk on large scales whose evolution results from balancing the scaling of a free string gas and the production of closed string loops by the interconnecting long strings~\cite{Vilenkin:1981kz, Vilenkin:1982ks, Vilenkin:1984ib, Vilenkin:2000jqa}. The network of global strings relaxes into an approximately scaling solution which is characterised by a correlation length $L$. From dimensional arguments, we expect that after some relaxation period, the energy density of a cosmic string scales as
\begin{equation}
    \rho_s = \xi(t) \frac{\mu_{\rm eff}(t)}{t^2} = \frac{\mu(t)}{L^2},
    \label{eq:axionicstring_simple}
\end{equation}
where $\xi$ is a dimensionless parameter describing the string stretching. It has been shown that the scaling $\rho_s \propto t^{-2}$ is attained in numerical simulations of {\it local} strings~\cite{Yamaguchi:1998gx, Gorghetto:2018myk} after an initial fast relaxation period, while logarithmic corrections to both $\xi$ and $\mu$ are expected for global strings due to their couplings to axions. The scaling regime described in Eq.~\eqref{eq:axionicstring_simple} represents the large-scale behaviour of the string network and holds when the string has completely stretched and modes smaller than the horizon at time $t$ do not contribute significantly to the total energy density~\cite{Vilenkin:1981kz}. In the standard scenario that makes up the radiation-dominated cosmology, global strings have not completely straighten, so that a contribution from short wavelengths to the total energy density arises as a logarithmic correction to a constant value of the linear mass density. For the case of the U(1) PQ symmetry with the potential given in Eq.~\eqref{eq:PQpotential}, the linear mass density resulting from the string Lagrangian reads~\cite{Vilenkin:1982ks, Davis:1986xc}
\be
	\mu_{\rm eff}(t) = \pi f_A^2\ln\left(f_A t\right).
	\label{eq:mass_length}
\ee
Once the linear mass density $\mu_{\rm eff}(t)$ has been defined, the evolution of the string network is sketched within a specific framework like the one-scale model in its original formulation~\cite{Albrecht:1989mk} or when the dependence on the velocity of the centre of mass of the string is taken into account~\cite{Martins:1996jp, Martins:2000cs, Martins:2003vd}. The latter approach is suitable when including the effects of friction due to interaction of the network of strings with surrounding matter and radiation~\cite{Vilenkin:1984ib, Martins:1995tg}, and has been recently applied to study the evolution of axionic strings~\cite{Fleury:2015aca, Martins:2018dqg}.

\subsection{Power loss from string loops}

Typically, a network of cosmic strings consists of horizon-length long strings as well as smaller string loops. The long strings carry a net conserved topological charge and are stable, while closed loops which do not have a net charge form by the intersection and breaking of long intercommuting strings to then oscillate and decay through the emission of axionic and radial modes, with a negligible contribution in gravitational waves. A key ingredient in the evolution of the cosmic string networks is the interaction and the intersection between long strings during a collision, which leads to the production of cosmic string loops. The process is parametrised by a ``chopping'' phenomenological quantity $c$ which parametrises the efficiency of the loop-production mechanism. Numerical simulations of string networks in either matter- or radiation-dominated cosmologies indicate $c = 0.23 \pm 0.04$~\cite{Martins:2000cs}. In this framework, the conversion of the long string network energy density into loops is described as
\be
	\frac{d\rho_{\rm loop}}{dt}= \xi \frac{\mu_{\rm eff}(t)}{t^3}.
	\label{eq:loop_conversion}
\ee
To describe the power loss of the network into radiation, we consider the dissipation of the energy of a closed loop $E_{\rm loop} = \mu_{\rm eff} \ell$ with length $\ell$ into axions and gravitational waves~\cite{Battye:1993jv, Battye:1994qa, Martins:1995tg},
\be
	P_{\rm loop} = \frac{dE_{\rm loop}}{dt} = \kappa\mu_{\rm eff} + \gamma_{\rm GW}G\mu_{\rm eff}^2,
	\label{eq:decayrateloop}
\ee
where $\gamma_{\rm GW} \approx 65$ and $\kappa \approx 0.15$ are \textit{dimensionless} quantities~\cite{Vachaspati:1986cc, Sakellariadou:1990ne, Sakellariadou:1991sd} computed at a fixed value of the string velocity $\langle v^2\rangle \approx 0.5$. Since the ratio of the power loss in gravity waves and axions is of the order of $(f_A/\MP)^2$, which is a quantity much smaller than unity for any QCD axion model, we can neglect the contribution from the dissipation of the string network into GW when discussing the radiation of axions from strings. Using Eq.~\eqref{eq:decayrateloop}, the shrinking of a loop with initial size $\ell_i$ is described by the expression
\be
	\frac{d\ell}{dt} = \kappa - \frac{\ell}{\mu_{\rm eff}}\frac{d\mu_{\rm eff}}{dt},
\ee
where $\ell = \ell(t, \ell_i)$ is the size at time $t$. If we neglect the small time dependence of $\mu_{\rm eff}(t)$ in the computations, the shrinking rate of the loop size with time is described by a linear function,
\be
	\ell(t) = \ell(t_i) - \kappa (t-t_i),
	\label{eq:shrinking}
\ee
in which the first term is the initial loop size and the second term describes the shrinking of the loop when emitting axions. Although in principle, the length of loops formed could be ranging at any size, numerical simulations~\cite{Battye:1994au, Vanchurin:2005pa, Martins:2005es, Ringeval:2005kr, Olum:2006ix, BlancoPillado:2011dq, Blanco-Pillado:2017oxo} show that the initial length of the large loop at its formation tracks the time of formation as $\ell(t_i) =\alpha t_i$, where $\alpha$ is an approximately constant loop size parameter which gives the fraction of the horizon size at which loops predominantly form. Recent simulations find that around 10\% of the energy released by the network of long string is released into ``large'' loops of size $\ell(t_i) =\alpha t_i$, while the remaining energy is lost into highly-boosted smaller loops. For this reason, a monochromatic loop spectrum is usually assumed~\cite{Kibble:1984hp, Cui:2008bd}, described by a loop formation rate $r_\ell(\ell_i, t_i)$ which is defined so that each loop of size $\ell$ contributes a factor $\rho_\ell(\ell,t)d\ell = \mu_{\rm eff}(t) \ell t_\ell(\ell, t)d\ell dt$ to the energy density in the time interval $dt$. The formation of loops from the long string condensate is described by
\be
	\frac{\rho_{\rm loop}}{dt} = \int_0^{+\infty} d\ell\, \mu_{\rm eff}(t) \,\ell\, r_\ell(\ell, t).
	\label{eq:loop_conversion1}
\ee
Here, we set $\alpha = 0.1$.

\subsection{Evolution of the string network}

The loop of initial size $\ell_i$ shrinks and disappears by the time $t_f$ defined as
\be
	t_f = t_i\,\(\frac{\alpha}{\kappa} + 1\).
	\label{eq:define_tf}
\ee
At any subsequent time $t$ after the formation of the loop, the energy density in radiation (either axions, GW, or other massless modes) from loops is given by
\be
	\rho_{\rm rad}^{\rm loop} = \int_0^{+\infty} \!\!d\ell_i \int_{t_{\rm PQ}}^t\!\! dt_i\(\!\frac{a(t_i)}{a(t)}\!\)^3 \int_{t_i}^{t_f} d\tau P_{\rm rad}(\tau)r_\ell(\ell_i, t_i),
	\label{eq:axionsenergydensityfromloop}
\ee
where we have introduced the power spectrum $P_{\rm rad}(\tau)$ for the decay of the loop energy into the radiation mode considered. The string network evolves from the time $t_{\rm PQ}$ at which the spontaneous symmetry occurs until the axion field acquires a mass at $t_{\rm osc}$, at which point the domain walls dissipate the string network which quickly decays. During this period, loops form at time $t_i$ and evolve emitting cold axions in their spectra.

We model the equation of state of the string network as $p_{\rm str} = \gamma \rho_{\rm str}$, with $-1/3 \leq \gamma \leq 1/3$~\cite{Vachaspati:1986cc}. In general, the equation of state depends on the mean-squared velocity of the long strings $\langle v^2\rangle$ as $\gamma = \(2\langle v^2\rangle - 1\)/3$, so that for $\langle v^2\rangle \approx 1$ (ultra-relativistic strings) we obtain the equation of state corresponding to radiation, while in the case of slow-moving strings $\langle v^2\rangle \approx 0$ the strings stretch due to the cosmological expansion of the Universe so that the come to dominate the universe. Numerical simulations of long strings~\cite{Bennett:1987vf} show that $\langle v^2\rangle \approx 0.43$ during a radiation-dominated stage. The evolution of a free, non-intercommuting string is then described by the kinetic equation
\begin{equation}
    \frac{d\rho_{\rm free}}{dt} + 3H(1+\gamma)\rho_{\rm free} = 0,
    \label{eq:freestrings}
\end{equation}
with solution $\rho_{\rm free} \propto a^{-3(1+\gamma)}$. Assuming that the scaling regime in Eq.~\eqref{eq:axionicstring_simple} is attained and the energy density of the string scales as $\rho_{\rm free} \propto t^{-2}$ for a constant value of the string tension, we obtain that the evolution of the string network demands $\gamma = w_{\rm eff}$, that is, the pressure of the string network has to match that of the NSC fluid that dominates the universe to maintain the scaling describing the free-string gas in Eq.~\eqref{eq:freestrings}. In other words, the Hubble equation $H^2 \propto \rho_\phi$ enforces that the NSC fluid also scales with the same equation of state as the string network so that $\rho_\phi \propto t^{-2}$. For $\gamma = -1/3$ we obtain the evolution $\dot\rho_{\rm free} + 2H\rho_{\rm free} = 0$, where the factor of two multiplying the Hubble expansion rate describes the combination of the volume dilution at the rate $3H\rho_{\rm free}$, and the stretching at the rate $-\gamma H\rho_{\rm free}$ so that the energy per unit length of the string remains constant as it gets stretched by the Hubble expansion. In this case, the formula in Eq.~\eqref{eq:freestrings} predicts the energy density dilution $\rho_{\rm free} \propto a^{-2}$, so that a network of free strings with $\gamma = -1/3$ would quickly come to dominate the expansion rate of the universe if they do not dissipate into Goldstone bosons (axions) or gravitational waves. Here, we set $\langle v^2\rangle = 0.5$ for both long strings and loops, which is consistent with the choice $\gamma = 0$.

\subsection{Axion energy density from string loops}

The picture modifies when including the emission of axions and other relativistic and radial modes like GW from the decay of the long string network, which is obtained by extending Eq.~\eqref{eq:freestrings} to include an energy density rate $\Gamma_{\rm str \to A}$~\cite{Harari:1987ht}. For this, we define the energy density of the free string gas as the solution to Eq.~\eqref{eq:freestrings} and with initial energy density equal to that of the string network in Eq.~\eqref{eq:axionicstring_simple},
\be
	\rho_{\rm free}(t) = \xi \frac{\mu_{\rm eff}(t)}{t_{\rm PQ}^2}\(\frac{t}{t_{\rm PQ}}\)^{-2\frac{1+\gamma}{1+w_{\rm eff}}},
\ee
so that the energy lost in the emission of axions and GW per unit time $\Gamma_{\rm str \to A}$ is the difference in the rates at which the energy densities in the free string gas and the string network change,
\bea
	\Gamma_{\rm str \to A}(t) &\equiv& \dot\rho_{\rm free}(t) - \dot\rho_s(t) = \nonumber \\
	&=& 3H(w_{\rm eff} - \gamma)\rho_s - \frac{\dot \mu_{\rm eff}}{\mu_{\rm eff}}\rho_s.
	\label{eq:globalstring}
\eea
This expression generalises the result in Eq.~\eqref{eq:loop_conversion} to NSC scenarios and to a time-dependent $\mu_{\rm eff}(t)$. For $w_{\rm eff} = \gamma$, the right-hand side of Eq.~\eqref{eq:globalstring} is zero, so that there is no transfer of energy between the string network and the axion population. In this limiting case, the evolution of the global string network matches that of a free string gas $\rho_s$ similarly to the dissipation-less case described in Eq.~\eqref{eq:freestrings} for $\Gamma_{\rm str \to A} = 0$, while the number density of axions remains unchanged and scales with the inverse volume. When $w_{\rm eff} < \gamma$, the right-hand side in Eq.~\eqref{eq:globalstring} changes sign and the axion condensate would feed energy into the cosmic string network. For these reasons, we set the number density in axions from strings in Eq.~\eqref{eq:numberdensityaxions1} equal to zero for $w_{\rm eff} \leq \gamma$.

The energy density of the radiated axions follows the evolution $\rho_A + 4H\rho_A = \Gamma_{\rm str \to A}$, to which it corresponds the energy density in Eq.~\eqref{eq:axionsenergydensityfromloop} and the number density of axions~\cite{Gorghetto:2018myk}
\be
	n_A^{\rm str} = \int^t\,dt'\frac{\Gamma_{\rm str \to A}(t')}{H(t')}\(\frac{a(t')}{a(t)}\)^3\,\int\frac{dk}{k}\,F(k),
	\label{eq:numberdensityaxions}
\ee
where the spectral energy density is defined in terms of a spectral index $q>1$ that describes a power spectrum ranging over all modes from $k \approx 1/\ell(t_i) \approx H(t_i)/\alpha$ to infinity. Demanding that the spectrum is normalised over the interval given results in~\cite{Battye:1993jv, Battye:1994qa, Gorghetto:2018myk}
\be
	F(k) = \frac{q-1}{\alpha^{q-1}}\(\frac{k}{H}\)^{-q}.
	\label{eq_axionspectrum}
\ee
In order to correctly compute the integrals in Eq.~\eqref{eq:numberdensityaxions}, we have to express the ratio $a(t')/a(t)$ in two different scenarios. If the axion acquires a mass at a temperature $T_{\rm osc} \lesssim \TRH$, we model the expansion rate appearing in the definition of $t_i$ as
\be
	\frac{a(t')}{a(t_{\rm osc})} = \begin{cases}
	\(\frac{t'}{t_{\rm RH}}\)^\beta\(\frac{t_{\rm RH}}{t_{\rm osc}}\)^{1/2}, & \hbox{for $t' < t_{\rm RH}$},\\
	\(\frac{t'}{t_{\rm osc}}\)^{1/2}, & \hbox{for $t' \geq t_{\rm RH}$},\\	
	\end{cases}
	\label{eq:scalefactorNSC}
\ee
where $t_{\rm RH}$ is defined so that $H(t_{\rm RH}) \equiv H_{\rm RH}$.  If instead we are in the case in which the axion field experiments only the modified cosmology $T_{\rm osc} \gtrsim \TRH$, we obtain $a(t')/a(t_{\rm osc}) = (t'/t_{\rm osc})^\beta$.

\begin{figure}[tb]
\begin{center}
	\includegraphics[width=\linewidth]{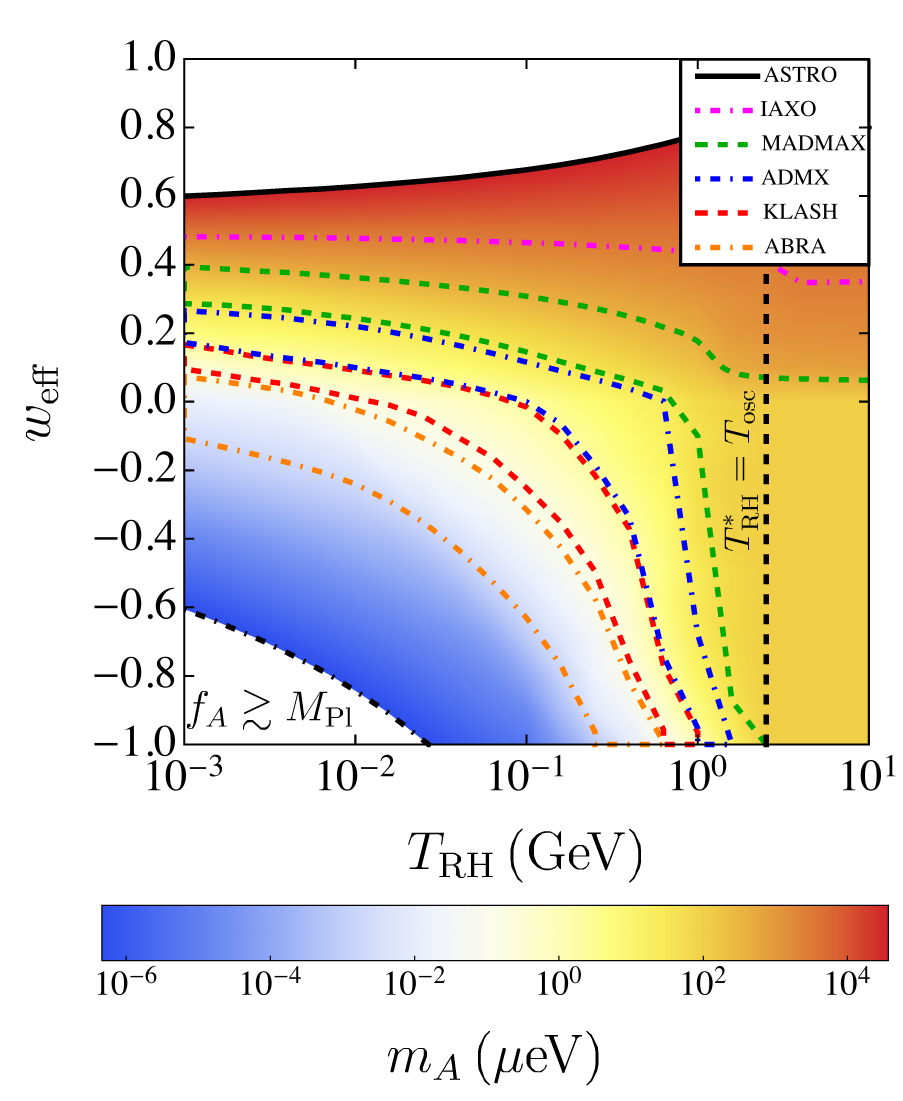}
	\caption{The value of the axion mass (in ${\rm \mu eV})$ that yields the observed dark matter abundance for different cosmological scenarios. The vertical axis reports the effective equation of state $w_{\rm eff}$, while the horizontal axis is the reheat temperature. The dashed vertical black line marks the reheat temperature $\TRH^*$ that matches the temperature at which the axion acquires a mass in the standard scenario, $T_{\rm RH}^* = T_{\rm osc}$. We also show the sensitivity that is expected to be reached by ABRACADABRA (``ABRA'', orange dot-dashed line), KLASH (red dashed line), ADMX (blue dot-dashed line), MADMAX (green dashed line), and IAXO (magenta dot-dashed line). The black dashed line marks the region beyond which the axion decay constant violates the weak gravity conjecture, $f_A \gtrsim \MP$, while the black solid line ``ASTRO'' gives the bound on $m_A \gtrsim 15{\rm meV}$ excluded by astrophysical considerations~\cite{RAFFELT1986402, Raffelt2008, Viaux:2013lha, Giannotti:2017hny}.}
	\label{fig:RecapFigureStrings}
\end{center}
\end{figure}

The integration of Eq.~\eqref{eq:numberdensityaxions} with the spectrum in Eq.~\eqref{eq_axionspectrum} leads to the result that most of the axions are radiated by loops at times right before $t_{\rm osc}$ and domain walls dissipate the network. The computation matches the estimation for the decay of the string network at the time of domain wall formation, as discussed in Refs.~\cite{Davis:1986xc, Battye:1993jv}. The number density of axions at time $t_{\rm osc}$ from Eq.~\eqref{eq:numberdensityaxions} results from a steeply falling integrand function of $t$, so that the dominant contribution comes from loops originating nearly instantaneously at values $t_i \sim t_{\rm osc}$~\cite{Davis:1986xc, Battye:1993jv, Battye:1994au}. If we refer the value of $n_A^{\rm str}$ to the contribution from the misalignment mechanism,
\begin{equation}
	n_A^{\rm mis}(t_{\rm osc}) \approx \frac{1}{2}m_A(T_{\rm osc})f_A^2\langle\theta_i^2\rangle \approx \frac{f_A^2\,\langle\theta_i^2\rangle}{(1+w_{\rm eff})t_{\rm osc}},
\end{equation}
Neglecting for the moment the time dependence in $\mu_{\rm eff}$, the contribution to the axion number density from the string network decay is
\bea
	\frac{n_A^{\rm str}}{n_A^{\rm mis}} &\approx& \frac{3\xi \alpha}{\langle\theta_i^2\rangle}\frac{\mu_{\rm eff}}{f_A^2}\frac{(w_{\rm eff} - \gamma) (1 + w_{\rm eff})^2}{1-w_{\rm eff}}\(1 \!-\! \frac{1}{q}\)\Upsilon\!\(\frac{t_{\rm RH}}{t_{\rm osc}}\),\nonumber\\
	\Upsilon(x) &=& \begin{cases}
	1, & \hbox{$x \geq 1$},\\
	1 - \frac{1 - 3 w_{\rm eff}}{1 + w_{\rm eff}}\(\sqrt{x}-1\), & \hbox{$x < 1$},\\
	\end{cases}
	\label{eq:numberdensityaxions1}
\eea
The function $\Upsilon(x)$ appearing in Eq.~\eqref{eq:numberdensityaxions1} takes into account the non-negligible extra contribution from strings in a NSC, for the period $t < t_{\rm osc}$. The string contribution is thus suppressed when the equation of state is milder than the standard cosmology and enhanced for a stiff $w_{\rm eff} > 1/3$, consistently with previous findings. In Fig.~\ref{fig:RecapFigureStrings} we show the effect of adding the extra energy density in strings to obtain the value of the DM axion mass. We have integrated Eq.~\eqref{eq:numberdensityaxions} numerically by including the contribution from a time-varying linear mass density and the scale factor obtained from the set of Boltzmann Eqs.~\eqref{eq:Kinetic_Rad}-\eqref{eq:Kinetic_psi}. While the region $w_{\rm eff} < \gamma$ remains unmodified, the effects of adding the string contribution alter the total energy density of axions, generally leading to higher values of the DM axion mass.

The results obtained in this Section have been derived for a power spectrum of index $q > 1$, which assumes that axions are radiated away on a timescale comparable to the Hubble time. It is worth noting that an alternative result for a harder spectral index $q=1$ has been presented in the literature on axion cosmology~\cite{Harari:1987ht, Hagmann:1990mj, Chang:1998tb}, starting from the assumption that strings efficiently shrink emitting all of their energy at once, leading to a flat power spectrum per logarithmic interval with an infrared cutoff at the wave mode $k \approx H$ and a ultraviolet cutoff at $k = f_A$. An excellent comparison of these different scenarios has been recently given in Ref.~\cite{Gorghetto:2018myk}. The use of the harder spectrum with $q=1$ generically leads to results for the string contributions which are smaller by a factor $\log(f_A t_{\rm osc}) \approx 70$. Clearly, in the case $q=1$ the results in Eq.~\eqref{eq:numberdensityaxions1} cannot be applied, and the integrated spectrum has to be re-computed from Eq.~\eqref{eq:numberdensityaxions}. The results expressed in Fig.~\ref{fig:RecapFigureStrings} can thus be interpreted as the situation in which the contribution from axionic strings is the largest, both because we have set a value $\xi = 10$ and because we have used the spectral index $q > 1$ given in Refs.~\cite{Davis:1985pt, Davis:1986xc, Battye:1993jv, Battye:1994au}, which leads to an additional enhancement by a factor or order $\approx 70$. In this view, the difference between the results in Fig.~\ref{fig:axionmass_nostrings} and in Fig.~\ref{fig:RecapFigureStrings} can be interpreted as the uncertainties in the experimental reach of future detectors due to topological defects.

\section{Gravitational Waves from Axionic String Loops} \label{Gravitational Waves from Axionic String Loops}

\subsection{Relic energy density of GWs from a string network}

In this Section, we compute the stochastic background GW emitted from axionic string loops. We consider a global string network that maintains its scaling regime either by radiating into axion or by the generation of global string loops. The contribution from axionic string radiation into GWs is generally suppressed by a factor $\(f_A/\MP\)^2$ with respect to the power injected into axions. The fraction of the critical energy density that is released into the GW spectrum per unit logarithmic interval of frequency is expressed by
\be
	\Omega_{\rm GW}(t, f) = \frac{1}{\rho_c(t)}\frac{d\rho_{\rm GW}}{d\ln k}.
\ee
The evolution of the string loop is described by Eq.~\eqref{eq:shrinking} where, contrarily to the previous literature, we have taken $\kappa \sim 0.15$ which characterises the predominant energy loss into axions, instead of using the value $\kappa \sim \gamma_{\rm GW}G\mu_{\rm eff}$ which would describe the predominant release of energy into gravity wave modes. As discussed in the previous Section, string loops oscillate at the modal frequencies $f_{\rm emit} = 2n/\ell(t')$ at the emission time $t'$ and for any integer $n$, decaying and emitting axions and GWs in the process according to the power loss in Eq.~\eqref{eq:decayrateloop}. The emitted modes redshifts with the expansion of the universe, so that the frequency $f$ observed at a later time $t > t'$ is
\be
	f = \frac{2n}{\alpha t_i - \kappa(t' - t_i)}\frac{a(t')}{a(t)},
	\label{eq:frequency_redshift}
\ee
where we have used Eq.~\eqref{eq:shrinking} to express the loop size $\ell(t')$. Inverting Eq.~\eqref{eq:frequency_redshift} gives the time $t_i$ at which the loop has formed as a function of the frequency,
\be
	t_i = \frac{1}{\alpha + \kappa} \(\frac{a(t')}{a(t)}\,\frac{2 n}{f} + \kappa t'\).
	\label{eq:frequency_redshift1}
\ee
The spectrum of radiated axions or GW is modelled by the emission from the loop shrinking with some power spectrum $P_n$ as
\be
	P_{\rm GW}(t, f') = \gamma_{\rm GW} \sum_{n=1}^{+\infty}\frac{\ell_n}{f'} {\bf n}_\ell(\ell, t)\,P_n,
	\label{eq:powerspectrum}
\ee
where ${\bf n}_\ell(\ell, t)$ is the number density of loops of size $\ell_i$ in the interval $d\ell$, which receives contributions from all pre-existing loops that have shrunk to the physical lengths $\ell(t')$ at time $t'$ as~\cite{Kibble:1984hp, Vilenkin:1984ib, Sousa:2013aaa, Sousa:2016ggw}
\be
	{\bf n}_\ell(\ell, t) = \frac{\xi}{\alpha}\(\frac{a(t_i)}{a(t')}\)^3\,\frac{1}{t_i^4}.
	\label{eq:define_r}
\ee
We have fixed the pre-factor in Eq.~\eqref{eq:define_r} by matching the condition in Eq.~\eqref{eq:loop_conversion}, linking the energy transfer between long strings and loops, with the definition in Eq.~\eqref{eq:loop_conversion1} for a monochromatic spectrum. Notice that the results in Ref.~\cite{Kibble:1984hp} offer an analytic formula for the number density of loops which has been tested in numerical simulations in matter-dominated~\cite{Yamaguchi:1999dy} and radiation-dominated cosmologies~\cite{Battye:1994au}. However, we have checked that we get to the same qualitative result by using Eq.~\eqref{eq:define_r}. In particular, Eq.~\eqref{eq:define_r} gives the same dependence on the NSC history ${\bf n}_\ell(\ell, t) \propto t_i^{3\beta-4}$ as what has been obtained in Ref.~\cite{Kibble:1984hp}, as we explicitly show in Eq.~\eqref{eq:GWmode1} below. In Eq.~\eqref{eq:powerspectrum}, we have decomposed the total power emitted by a slowly moving loop in Eq.~\eqref{eq:OmegaGW} as $P = \sum_n P_n$, where the spectral mode of frequency $f_n$ is reasonably described by the distribution of power per mode of emission~\cite{Allen:1990tv, Blanco-Pillado:2013qja, Blanco-Pillado:2017oxo}
\be
	P_n = \frac{\gamma_{\rm GW}\,n^{-4/3}}{\mathcal{N}},
	\label{eq:spectral_power}
\ee
where $\mathcal{N} = \sum_n n^{-4/3} \approx 3.60$. The power spectral index $q=4/3$ is characteristic of the emission of GW modes from the decay of cuspy loops, as suggested by numerical simulations~\cite{Allen:1990tv}~\footnote{Analytic results~\cite{Garfinkle:1987yw} have pointed out a harder spectrum $q=2$.}. Since the sum in the expression for the total emission converges slowly, higher emission modes significantly contribute to the total power. Finally, the fractional contribution to the energy density from the gravitational wave spectrum is given by
\be
	\Omega_{\rm GW}(t, f) = \sum_n P_n\, \Omega_{\rm GW}^{(n)}(t, f),
	\label{eq:OmegaGW}
\ee
where the contribution from the mode $n$ reads~\cite{Cui:2018rwi}
\be
	\Omega_{\rm GW}^{(n)}(t, f) \!=\! \frac{1}{\rho_c(t)} \frac{2n}{f}\frac{\xi}{\alpha} \int_{t_s}^t dt' \frac{G\mu_{\rm eff}^2(t')}{t_i^4}\(\frac{a(t')}{a(t)}\)^5\(\frac{a(t_i)}{a(t')}\)^3.
	\label{eq:GWmode}
\ee
To assure ourselves that we have control over the numerics involved in solving Eq.~\eqref{eq:GWmode}, we have first reproduced the same quantitative results as what has recently obtained in Ref.~\cite{Cui:2018rwi} for the case of a global string network lasting until present time $t = t_0$ and for string loops that predominantly emit into gravity wave modes, for which $\kappa = \gamma_{\rm GW} G\mu_{\rm eff}$ (not shown).

\subsection{GW spectrum from a network of global strings} \label{sec:GW spectrum from a network of global strings}

\begin{figure}[tb]
\begin{center}
	\includegraphics[width=\linewidth]{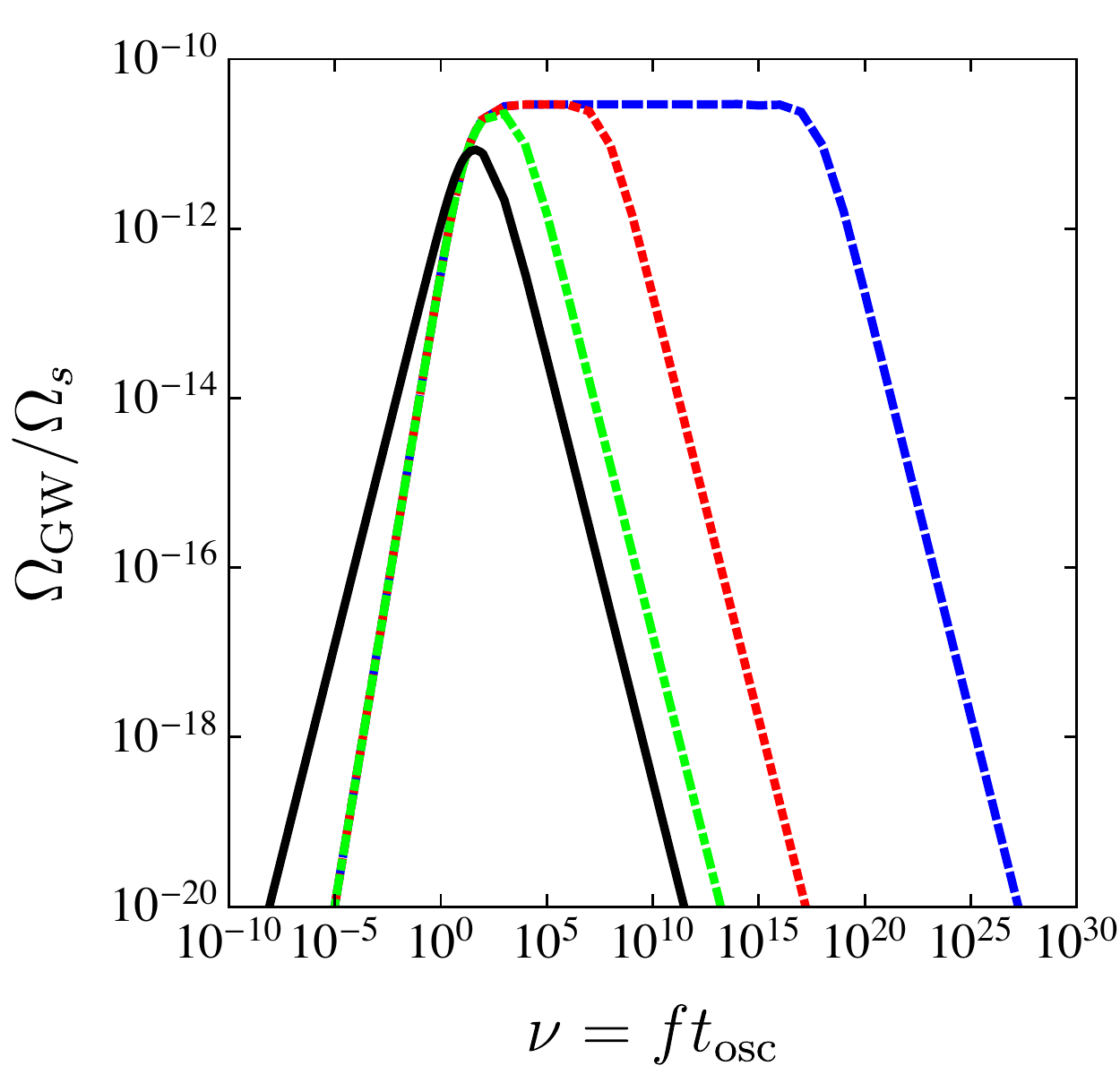}
	\caption{The fractional contribution to the GW energy density $\Omega_{\rm GW}$ in Eq.~\eqref{eq:GWmode} in units of $\Omega_s(t_{\rm osc})$, as a function of the quantity $\nu = f t_{\rm osc}$, assuming that an early matter-dominated stage has taken place to time $t_{\rm RH} = x_{\rm RH}t_{\rm osc}$. We have plotted the results for $x_{\rm RH} = 10^{-32}$ (blue dashed line), $x_{\rm RH} = 10^{-12}$ (red dotted line), $x_{\rm RH} = 10^{-4}$ (green dot-dashed line), and $x_{\rm RH} > 1$ (black solid line).}
	\label{fig:GW0Figure}
\end{center}
\end{figure}

We have tested the expression in Eq.~\eqref{eq:GWmode} in a cosmological model composed of an early NSC period with scale factor $a(t)\propto t^\beta$, followed by a period of radiation-domination. In this toy model, we neglect the time dependence in the linear mass density, so we assume that $\mu_{\rm eff}$ does not depend on time, and we set the time at which the symmetry breaking occurs as $t_s = 0$. In particular, in the case $T_{\rm osc} \gtrsim \TRH$, the expression in Eq.~\eqref{eq:GWmode} can be written in a simpler form which we show for illustration,
\bea
	\Omega_{\rm GW}^{(n)}(\nu) &=& \frac{\rho_s(t_{\rm osc})}{\rho_c(t_{\rm osc})} \frac{2n}{\nu}\frac{G\mu_{\rm eff}}{\alpha} \int_0^1 dx x^{2\beta}\,x_i^{3\beta-4},\\
	x_i &=& \frac{1}{\alpha + \kappa} \(\frac{2 n}{\nu}x^\beta + \kappa x\),
	\label{eq:GWmode1}
\eea
where $\nu = ft_{\rm osc}$ and the energy density in strings $\rho_s(t)$ has been given in Eq.~\eqref{eq:axionicstring_simple}. We assume that the toy model comprises an early MD scenario with $\beta = 2/3$, in which the universe transitions to the radiation-dominated stage at temperature $\TRH$. We show results in Fig.~\ref{fig:GW0Figure}, where we plot the contribution in GW to the total energy density, in units of $\Omega_s = \rho_s(t_{\rm osc})/\rho_c(t_{\rm osc})$, as a function of $\nu$ so that the plot we show is independent on the choice of $t_{\rm osc}$. For obtaining the results in Fig.~\ref{fig:GW0Figure}, we have fixed the constant $G\mu_{\rm eff} = 10^{-11}$ and $\kappa = 0.15$. The quantity $x_{\rm RH} = t_{\rm RH}/t_{\rm osc}$ parametrises the duration of the MD scenario from early time $t=0$ on. For $x_{\rm RH} \to 0$ (the blue dashed line in Fig.~\ref{fig:GW0Figure} has been plotted for $x_{\rm RH} = 10^{-32}$, the red dotted line for $x_{\rm RH} = 10^{-12}$), the toy model describes a string network decaying into GWs in a radiation-dominated background, while as $x_{\rm RH}$ approaches unity, the contribution from the early MD period becomes more significant (the green dot-dashed line in the figure has been plotted for $x_{\rm RH} = 10^{-4}$). For $x_{\rm RH} > 1$ (solid black line in Fig.~\ref{fig:GW0Figure}), the string network decays during MD cosmology and it is not sensitive on the details of the subsequent reheating. For $x_{\rm RH} = 0$, the string network releases GWs in a radiation-dominated cosmology where $\Omega_{\rm GW} \propto f^{-3/2}$ for $\nu \lesssim \bar \nu$ and $\Omega_{\rm GW}$ is constant for $\nu \gtrsim \bar \nu$, where $\bar \nu$ is the value of $\nu$ for which the two terms appearing in the brackets in Eq.~\eqref{eq:frequency_redshift1} match. The second kink at larger values of $\nu$ corresponds to the change in the dependence of $t_i$ due to the non-standard cosmological model, according to the change in the scale factor expressed in Eq.~\eqref{eq:scalefactorNSC}. Note, that the dependence for lower values of $\nu$ also modifies, consistently with the findings of Ref.~\cite{Cui:2018rwi}.

\subsection{Energy density of GWs from axionic strings}

We now turn our attention back to the case of the axionic string network, for which different forms of the expressions in Eqs.~\eqref{eq:frequency_redshift1} and~\eqref{eq:GWmode} are possible depending whether the axion field starts to oscillate in the NSC or in the standard epoch. In the scenario where $T_{\rm osc} \lesssim \TRH$, the expression for $a(t')/a(t_{\rm osc})$ is given by Eq.~\eqref{eq:scalefactorNSC} and the axion field remains frozen at its vacuum expectation value after symmetry restoration. Although additional strings are no longer generated, the existing ones will continue producing GWs after reentering the expanding horizon. This GW production stops when the Hubble rate becomes comparable to the mass of the field, see Eq.~\eqref{eq:axioneqmotion}. From this moment on, the axion field oscillates about its minimum and the strings network decays. If instead we are in the case in which the axion field experiments only the modified cosmology $T_{\rm osc} \gtrsim \TRH$, we obtain $a(t')/a(t_{\rm osc}) = (t'/t_{\rm osc})^\beta$. In this second scenario, the symmetry is restored and the axion field oscillates about its minimum before radiation domination, halting the production of GWs well before the radiation-dominated era begins.

Our treatment for studying GWs from a realistic axionic string network differs in two key aspects from the results used in Ref.~\cite{Cui:2018rwi}. First of all, we do not consider a string network that persists until present time, but we only focus on the axionic string network that decays at time $t_{\rm osc}$. Secondly, GWs are a subdominant mechanism of energy release, the main contribution from strings going into radiated axions as discussed in Sec.~\ref{sec:axion_strings} for which $\kappa \approx 0.15$. We elaborate the results in Eq.~\eqref{eq:OmegaGW} to account for the dissipation of the string network at $t_{\rm osc}$ and the subsequent redshift of the energy density as
\bea
	&&\Omega_{\rm GW}(t_0, f_0) = \frac{\rho_c(t_{\rm osc})}{\rho_c(t_0)}\(\!\frac{a(t_{\rm osc})}{a(t_0)}\!\)^4 \Omega_{\rm GW}(t_{\rm osc}, f) = \nonumber \\
	= &&\Omega_{\rm R}h^2 \!\(\frac{g_*(T_0)}{g_*(T_{\rm RH})}\)^{\frac{1}{3}}\!\(\frac{a(t_{\rm RH})}{a(t_{\rm osc})}\)^{3w_{\rm eff} - 1} \!\!\!\!\!\Omega_{\rm GW}(t_{\rm osc}, f),
	\label{eq:OmegaGWred}
\eea
where in the last step we have used the fact that $\rho_c(t) \propto t^{-2} \propto a(t)^{-2/\beta}$~\cite{Hiramatsu:2013qaa} and the redshift of the radiation energy density as $\rho_{\rm rad}(T) \propto g_*(T)T^4$. In Eq.~\eqref{eq:OmegaGWred}, the frequency $f_0 = f a(t_{\rm osc})/a_0$ accounts for the redshift of the peak wavelength.

\begin{figure}[tb]
\begin{center}
	\includegraphics[width=\linewidth]{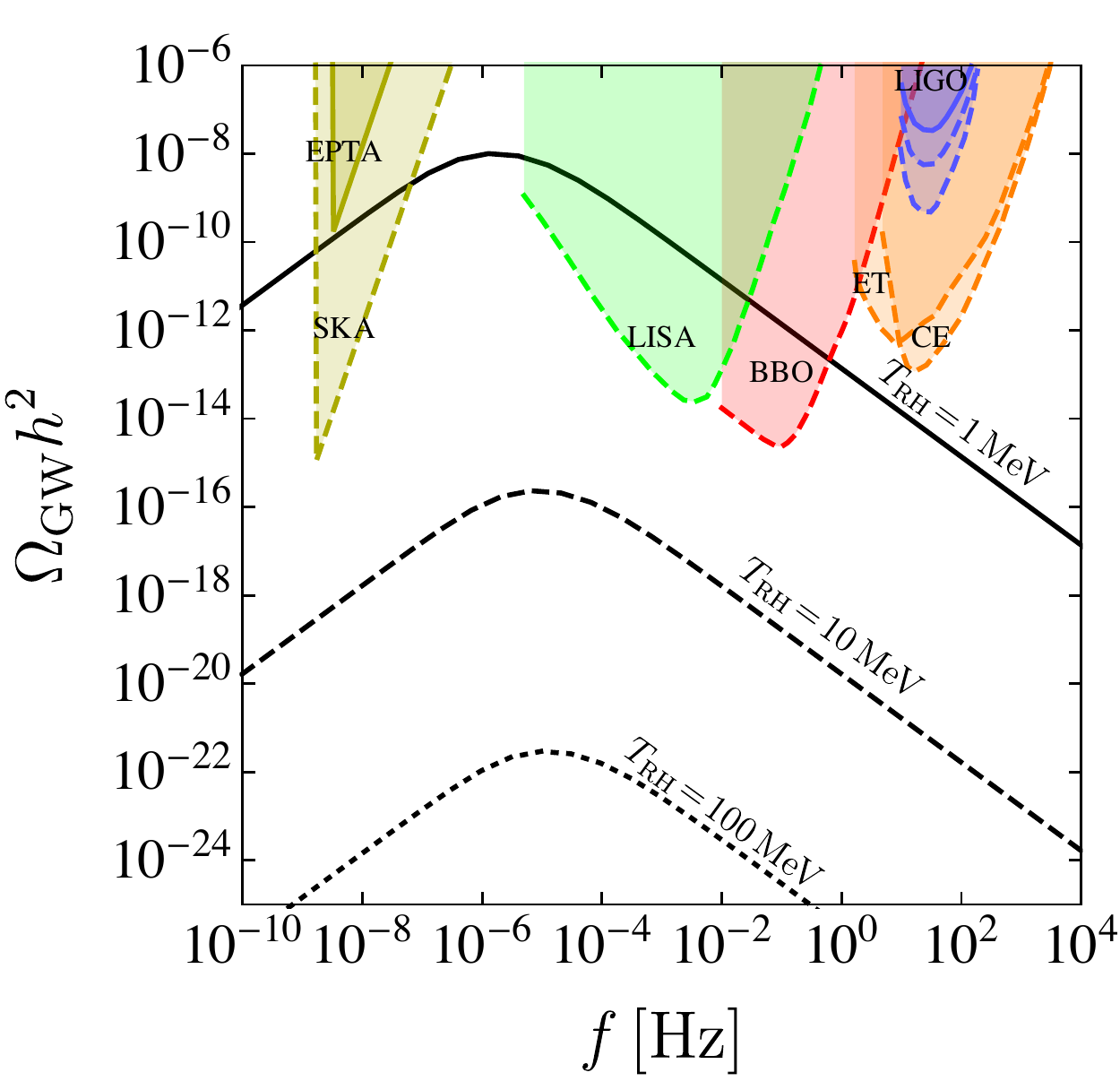}
	\caption{The fractional contribution to the GW energy density $\Omega_{\rm GW}$ to the critical density at present time, as a function of the frequency of the GW corrected for the redshift. We have plotted results for $\TRH = 1\,$MeV (solid black line), $\TRH = 10\,$MeV (dashed black line), $\TRH = 100\,$MeV (dotted black line). Also shown are the sensitivities of current (solid) and planned (dashed) experiments that are undergoing in detecting GWs, see the text fo additional information.}
	\label{fig:GWFigure}
\end{center}
\end{figure}

When the axion model is taken into account, the situation described in Fig.~\ref{fig:GW0Figure} changes dramatically, since the moment at which the axion field starts to oscillate and the value of $f_A$ that assures the axion to explain the totality of the dark matter depend on the value of $\TRH$. In particular, as we have discussed in the previous Sections, the DM axion mass is sensibly smaller when computed in early MD models with respect to its standard value, so that the corresponding value of $f_A$ increases by a factor $\mathcal{O}(10^2)$. Since the energy density in GWs in Eq.~\eqref{eq:GWmode} scales as $\propto f_At_{\rm osc}\propto \TRH f_a^{3/2}$, we expect a substantial enhancement if the modes are emitted during MD. In Fig.~\ref{fig:GWFigure} we show results for the present energy density in GW from axionic strings, in units of the present critical density, as a function of frequency once the redshift from $t_{\rm osc}$ has been taken into account. The shape of the GW spectrum in Fig.~\ref{fig:GWFigure} is substantially different from what has been obtained in the model depicted in Sec.~\ref{sec:GW spectrum from a network of global strings}. In fact, the shape of the GW spectra in Fig.~\ref{fig:GW0Figure} has been obtained by assuming a model in which the spectrum of the string network radiation explores different regimes. At low frequencies, the spectrum transitions from a raising spectrum to a plateau that corresponds to the change in the contribution from the two different terms in the square brackets of Eq.~\eqref{eq:frequency_redshift1}, while the second tilt at higher frequencies corresponds to the effects of the early matter-dominated stage. The two tilts in the power spectrum are not featured in the GW spectrum from axionic strings shown in Fig.~\ref{fig:GWFigure}, for which only one tilt is present and the plateau is missing. In fact, in this latter case the network of axionic strings decays during matter-domination and prior radiation-domination, so that the spectrum of GWs from axionic string loops vibrating in a MD cosmology does not show the flat plateau typical of the radiation-domination period~\cite{Cui:2018rwi}.

The model presented is barely within reach of future planned experiments like LISA~\cite{LISA:2017} and BBO~\cite{Yagi:2011wg} for extremely low values of the reheat temperature $\TRH$ as allowed by the BBN constraint. For higher values of $\TRH$, the energy density released in GWs from the string network drops to the value expected in the standard radiation-dominated cosmology. For completeness, Fig.~\ref{fig:GWFigure} also reports the current and future sensitivities of LIGO-VIRGO~\cite{TheLIGOScientific:2014jea, TheLIGOScientific:2016wyq, Abbott:2017mem}, and the projected sensitivities of both Einstein Telescope (ET)~\cite{Punturo:2010zz} and Cosmic Explorer (CE)~\cite{Evans:2016mbw}, which will probe a different frequency range with respect to the peak frequency in GWs from strings. Finally, we have shown the current limit placed by the European Pulsar Timing Array (EPTA)~\cite{vanHaasteren:2011ni}, and the expected sensitivity of the future Square Kilometre Array (SKA)~\cite{Janssen:2014dka}.

We briefly comment on an early period of a NSC dominated by a substance with a stiff equation of motion $w_{\rm eff} > 1/3$, bearing in mind the kination model as the archetype of such scenarios. In general, the primordial gravitational wave background for modes produced during such a NSC period would be blue tilted, as it can be predicted from computing Eq.~\eqref{eq:GWmode} in such a scenario. In particular, it has been suggested in previous literature~\cite{Bettoni:2018pbl, Cui:2018rwi} that a GW signal coming from loops oscillating during a kination stage would lead to a copious signal which would be easy to detect at higher frequencies, due to the dependence of the primordial gravitational wave background on the emitted frequency as $\Omega_{\rm GW} \propto f$. However, we note here that this is result does not hold for the case of an axionic string network, because the energy density in GWs strongly depends on the axion decay constant, $\Omega_{\rm GW} \propto f_At_{\rm osc}$, and the result that axion dark matter produced during kination possesses a larger mass thus a smaller value of $f_A$, see Fig.~\ref{fig:axionmass_nostrings}.

So far, we have extensively used the bound on the reheat temperature $\TRH \gtrsim 5\,$MeV obtained from requiring a successful BBN during radiation-domination. An similar constraint on the GW spectrum can be placed from the fact that GWs might excessively contribute to the effective number of relativistic degrees of freedom at BBN, as~\cite{Maggiore:1999vm, Caprini:2018mtu}
\be
	\mathcal{I} \equiv \int_{k_{\rm BBN}}^{k_{\rm end}} \Omega_{\rm GW}(k)h^2 \,d\ln k \lesssim 10^{-6},
	\label{eq:GWbound}
\ee
where $k_{\rm end}$ and $k_{\rm BBN}$ are the momenta associated to the horizon scale at the end of inflation and at BBN. Usually, this requirement is invoked to bound the duration of an early NSC phase with a stiff equation of state~\cite{Bettoni:2018pbl, Cui:2018rwi}. Here, we have checked that even in the most extreme case considered $\TRH = 1\,$MeV, the bound is respected since we obtain a value for the integral $\mathcal{I} \approx 10^{-7}$. A future improvement on the bound in Eq.~\eqref{eq:GWbound} could then be used to constrain $\TRH$ in an early MD scenario.

\section{Discussion and Conclusions}

In this paper, we have challenged the validity of the standard cosmological scenario in the context of the QCD axion as the dark matter particle. In the standard cosmological scenario, the post-inflationary universe is quickly reheated and it is described by a radiation-dominated epoch, down to temperatures of order the electron-volt at which the energy density in non-relativistic matter starts controlling the expansion rate right after matter-radiation equality. Matter domination continues until recent times, with solid indications for the present accelerated expansion driven by a cosmological constant. A large plethora of experiments support this view of the history of the universe, most notable large-scale structures, the CMB, and the successful predictions by BBN which so far allows us alone to rely on the standard cosmological model up to a temperature $\TRH \sim 5\,$MeV. No relic has been yet discovered to test earlier stages of the universe, although much work has been devoted onto such possibilities and their imprints into phenomenological expectations. The axion, both as a hot dark matter and as a cold dark matter component, has been studied as a possible probe of the pre-BBN epoch, and it has been shown that its properties like the present energy abundance, the velocity dispersion, and the size of the smallest structures formed strongly depend on the details of the cosmology at which the axion field acquires a mass. Since the axion field begins to oscillate at around the QCD phase transition, the axion is sensitive only to about three orders of magnitude in temperatures (from the MeV to the GeV scale) to the presence of a hypothetical modified cosmology, since the axion field configuration is frozen at higher temperatures.

Nevertheless, important results can be drawn if an additional component has modified the cosmological background within the range of interest for the QCD axion. In this paper, we have first reviewed the results obtained for the dark matter axion mass in a NSC, extending the previous results that only focused on MD and kination models to generic cosmological models parametrised by a set of coupled Boltzmann equations. We have linked these new results to the reach of future axion searches, since various new experiments have been proposed in the recent year to scan the value of the axion mass in different ranges of the mass spectrum. In the standard cosmology, and for a value of the initial misalignment angle of order one, the axion mass is expected to be in the range $m_A \approx (10 - 100)\,\mu$eV, the uncertainty arising from the computations involving the decay of the axionic string network. As we show in Fig.~\ref{fig:axionmass_nostrings}, the axion mass in the standard cosmology is within the present scope of MADMAX~\cite{TheMADMAXWorkingGroup:2016hpc}. The same computation performed in a NSC allows the DM axion mass to span all possible values, further motivating searches outside the range obtained in the standard cosmology and hinting as the axion as a possible probe of the pre-BBN epoch. We have summarised the results in Fig.~\ref{fig:axionmass_nostrings}, which shows the value of the DM axion mass as a function of the two parameters describing the NSC period (the effective equation of state $w_{\rm eff}$ and the reheat temperature).

We have then considered including the contribution to the axion energy density from the decay of the axionic string network, which is inevitably formed if the PQ symmetry is broken in the post-inflationary universe. Such a process requires dedicated numerical simulations in order to be tested and to deeply appreciate the physics modelling and the tuning of the parameters involved. No simulations of the axionic string network in background cosmologies other than radiation and matter are currently available, nevertheless we rely on the theoretical description of the axionic string decay to draw conclusions on this topic, as summarised in Fig.~\ref{fig:RecapFigureStrings}. We have shown that the inclusion of the string network sensibly modifies the energy density and the DM mass of the QCD axion, especially when a stiff equation of state for the NSC is included. This result has been interpreted as an enhanced efficiency in the string network to release energy density into the radiated axions, similarly to what has been shown for the case of a global string network that emits gravity wave modes~\cite{BlancoPillado:2011dq, Blanco-Pillado:2013qja, Blanco-Pillado:2017oxo, Cui:2018rwi}. When strings are included, the value of the DM axion mass might change by orders of magnitude with respect to the result obtained by the misalignment mechanism only, when varying $w_{\rm eff}$. A second difference in the two scenarios comes from the sensitivity of the contribution to the early NSC period in the case in which the axion field acquires a mass in the standard cosmology. In fact, the misalignment contribution is not sensitive to the early content of the universe because its dynamics is frozen during that period, while the string component acquires contributions from the whole history of the string network development as expressed in the integral form in Eq.~\eqref{eq:numberdensityaxions} (although the dominant contribution is generated at time $t_{\rm osc}$). The detection of a relatively large value of the axion mass could then yield clues for the production of strings in the early universe, due to the rapid increase of the axion mass with the string contribution. This is due to the accumulation of strings which radiate axions until $t_{\rm osc}$, with the largest contribution coming in the case of a stiff equation of state $w_{\rm eff} > 1/3$. We have set to zero any contribution coming from the string network in the case in which the equation of state describing the NSC is milder than the string equation of state, $w_{\rm eff} < \gamma$, due to considerations on the Boltzmann equations involved with the system.

Finally, we have checked whether gravity waves from the axionic string network might be relevant if emitted during a NSC. The emission of GWs during the evolution of an axionic string is usually neglected, due to its strong suppression in power with respect to the dominant emission in axions, and only becomes relevant when the axion acquires a mass and the string network decays. We have found that the GW emission can also be neglected in most modified scenarios, however an interesting result in the extreme case of an early MD epoch lasting until $\TRH \sim 5\,$MeV would yield to a signal that is barely detectable in future experiments, see Fig.~\ref{fig:GWFigure}. The feature of this spectrum is such that the primordial gravitational wave background is red tilted, with $\Omega_{\rm GW} \propto f^{-1}$, peaking at the frequency $f_{\rm peak} \sim 10^{-6}\,$Hz not currently expected to be probed by future generations of experiments. This enhancement in the GW signal from the early MD stage comes from the dependence $\Omega_{\rm GW} \propto f_At_{\rm osc}$ and from the larger value of $f_A$ in this scenario with respect to the standard result. As we show in Fig.~\ref{fig:RecapFigureStrings}, the largest values of $f_A$ comes in the region $w_{\rm eff} < 1/3$ like an early MD. In principle, the detection of an axion of mass $\mathcal{O}(10^{-8})\,$eV could come along with crossing evidences from probing the primordial GW wave spectrum. The main conclusion of this paper is that once the axion has been detected, its properties will also provide a guideline for understanding the cosmology of the early universe. Testing the axion theory could come along with the possibility of detecting a primordial gravitational wave relic, which would be possible in some favoured non-standard cosmological scenarios.

\begin{acknowledgments}
We thank Maurizio Giannotti, Enrico Nardi, Javier Redondo, and Sunny Vagnozzi for useful suggestions that improved the current form of the paper. LV acknowledges support by the Vetenskapsr\r{a}det (Swedish Research Council) through contract No. 638-2013-8993, the Oskar Klein Centre for Cosmoparticle Physics, and the kind hospitality of the INFN Laboratori Nazionali di Frascati and Barry University where part of this work was carried out.
\end{acknowledgments}

\appendix

\bibliography{axBib.bib}

\end{document}